\documentclass[aps,prd,a4paper,showpacs,nofootinbib]{revtex4-1}
\usepackage{amsmath}
\usepackage{bm}

\begin{document}
\title{Nonlocality, No-Signalling and Bell's Theorem investigated by Weyl's Conformal Differential Geometry.}
\author{Francesco De Martini}
\affiliation{Accademia Nazionale dei Lincei, via della Lungara 10, 00165 Roma, Italy}
\email{francesco.demartini@uniroma1.it}
\author{Enrico Santamato}
\affiliation{Dipartimento di Fisica, Universit\`a di Napoli Federico II, Compl. Univ. di Monte S. Angelo, Napoli, Italy}
\email{enrico.santamato@na.infn.it}
\begin{abstract}
The principles and methods of the Conformal Quantum Geometrodynamics (CQG)\
based on the Weyl's differential geometry are presented. The theory applied
to the case of the relativistic \textit{single} quantum spin $\frac{1}{2}$
leads a novel and unconventional derivation of Dirac's equation. The further
extension of the theory to the case of two spins $\frac{1}{2}$ in EPR
entangled state and to the related violation of Bell's inequalities leads,
by an exact albeit non relativistic analysis, to an insightful resolution of
all paradoxes implied by \textit{quantum nonlocality.}
\end{abstract}
\maketitle
\section{Introduction}\label{sec:intro}
Since the 1935 publication of the famous paper by Einstein -- Podolsky --
Rosen (EPR), the awkward coexistence within the quantum lexicon of the
contradictory terms \textquotedblleft \textit{locality}\textquotedblright\
and \textquotedblleft \textit{nonlocality}\textquotedblright\ as primary
attributes to quantum mechanics (QM)\ has been a cause of concern and
confusion within the debate over the foundations of this central branch of
modern Science~\cite{EPR35,Haag,doplicher74,doplicher10}. On the other hand, the
confirmation by today innumerable experiments, following the first one by
Alain Aspect and coworkers, of the paradoxical violation of the Bell
inequalities emphasizes the dramatic content of the dispute \cite%
{aspect82,Bell,Redhead}. By referring to the implications of Relativity with
the nonlocal EPR correlations the philosopher Tim Maudlin writes:
\textquotedblleft \textit{One way or another, God has played us a nasty
trick. The voice of Nature has always been faint, but in this case it speaks
of riddles and mumbles as well..." } \cite{Maudlin}. Indeed the violation of
the Bell inequalities realized by all these experimental test implied the
existence of quite \textquotedblleft mysterious\textquotedblright\ nonlocal
correlations linking the outcomes of the measurements carried out over two
spatially distant particles. Since "\textit{correlations cry out for
explanations}" according to J. S. Bell, these experimental results started,
about thirty years ago, a theoretical endeavour aimed at discovering the
inner dynamics underlying such an enigma \cite{Bell}. Moreover, it has been
recognized that, since any transfer of information through the EPR\
correlations is forbidden by special relativity, a \textquotedblleft
no-signalling theorem\textquotedblright\ must hold. Recently, even this
theorem was verified experimentally\cite{deangelis07}.

Aimed at a clarification of such a fundamental, albeit intriguing
paradox the present Article is intended to tackle the EPR scheme by a very
general, insightful perspective. Driven by an accurate reconsideration of
all natural symmetries affecting the dynamics of particles, our theoretical
analysis is indeed based on the well known standard linear quantum theory.
Very remarkably, this linear theory has recently been found to be rooted in
the Hermann Weyl's conformal geometric invariance properties affecting the
very structure of all physical laws \cite{weyl,santamato85}. This concept
was well expressed by P.A.M.Dirac in a 1973 seminal paper\cite{dirac73}: "%
\textit{There is a strong reason in support of Weyl's theory. It appears as
one of the fundamental principles of Nature that the equations expressing
basic laws should be invariant under the widest possible group of
transformations. The confidence that one feels in Einstein }[general
relativity]\textit{\ theory arises because its equations are invariant under
the wide group of transformations of curvilinear coordinates in Riemannian
space...The passage to Weyl's geometry is a further step in the direction of
widening the group of transformations underlying the physical laws. One has
to consider transformations of gauge as well as transformations of
curvilinear coordinates and one has to take one's physical laws to be
invariant under all these transformations, which impose stringent conditions
on them....}". These stringent conditions indeed express the
conformal-covariance (co-covariance, or gauge-covariance) of all physical
laws, including the ones belonging to electromagnetism and to the standard
quantum dynamics, as we shall see in the present Article. According to the
Weyl's \textit{conformal} differential geometry, the formal expression of
all physical laws can be expressed in different "gauges" which are related
by a \ \textit{conformal mapping }preserving the\textit{\ }angles between
vectors. This theoretical approach, today well known in the domain of modern
General Relativity and Cosmology ~\cite{hochberg91,Lord,Hehl} has never been
consistently applied in the past to the analysis of a wide class of low
energy quantum phenomena, including atomic physics\cite{santamato85}.
Indeed, the correct application of gauge-covariance (or unit-covariance) to
quantum phenomena implies very subtle considerations that were overlooked by
Einstein, and later by Weyl himself, at the time in which this elegant
abelian gauge theory was first\ proposed (1918) \cite{weyl,Quigg}.\ These
considerations imply in the first place the correct choice \textit{by
definition }of several units in terms of which all physical quantities are
measured: these units must be mutually \textit{independent}\ in the sense
that a dimensionless unit cannot be constructed with them. It is
conventional in relativistic quantum theory to take $c$, $\hbar$ and $%
m_{e}$ (the electron mass) to be "\textit{constant by definition".} Other
gauges, e.g. by replacing $m_{e}$ with the gravitational constant $G$, lead
in general to different theories which are mutually connected by conformal
mapping \cite{Lord}.

Restricting ourselves to the main topic of present Article, we believe that,
since the Weyl's conformal gauge symmetry reflects very essential properties
of Nature, it must be rooted in the inner structure of any sensible,
complete quantum theory. In the present Article this general theory will be
referred to as "\textit{Conformal Quantum Geometrodynamics}" (CQG). This one
reduces, in a particular gauge, to the well known standard linear quantum
theory living in Hilbert spaces. In Section 2 the principles and methods of
CQG are presented extensively. There the fundamental equations of linear
quantum dynamics for particles, i.e. the Schr\={o}dinger's and the Dirac's
equations, are derived by an exact variational calculus, i.e. with no
approximations\cite{santamato13}. In Section 3 the CQG\ theory deals with
one particles with any spin and then is extended to the case of one and two
particles with spin-%
$\frac12$%
. Section 4 deals with the key topic of the present Article, the EPR\
scheme, i.e. the nonlocal correlations of two equal \textit{entangled}
\textit{particles} with spin-%
$\frac12$%
. The measurement by two remotely distant stations (Alice $(A)$ and Bob $(B)$%
) of the spin of two equal particles ($A$ and $B$)\ emerging from two
\textit{Stern-Gerlach Analyzers} (SGA) is thorougly analyzed by a
nonrelativistic approach in the CQG\ framework, without approximations.
There it is shown that CQG theory indeed naturally violates the Bell's
inequalities without making recourse to any additional nonlocality
assumption. In addition, in Section 5 the "\textit{no-signalling" }process,
i.e.the impossibility of mutual exchange of useful information between $A$
and $B$ for the same two spin-%
$\frac12$
particle system is also found to be a very natural consequence of CQG. A
brief discussion on the perspective of our results in the context of modern
Physics, is contained in Section~\ref{sec:conclusions}.
\section{The Weyl's Conformal Geometrodynamics}
We consider a mechanical system described by $n$ generalized coordinates $%
q^{i}$ $(i=1,\dots ,n)$ spanning the configuration space $V_{n}$. The system
defines a metric tensor $g_{ij}(q)$ in $V_{n}$, for example by its kinetic
energy. However, even if the metric is prescribed, the geometrical structure
of $V_{n}$ is fully determined only after the parallel transport law for
vectors is also given. We assume an affine transport law given by the
connection fields $\Gamma _{jk}^{i}(q)$ with zero torsion, i.e. $\Gamma
_{jk}^{i}-\Gamma _{kj}^{i}=0$. The connection fields $\Gamma _{jk}^{i}(q)$
and their derivatives define in $V_{n}$ a curvature tensor $R_{jkl}^{i}$
and, together with the metric tensor, a scalar curvature field $%
R(q)=g^{ij}R_{ikj}^{k}$.\newline
We introduce the multiple-integral variational principle
\begin{equation}\label{eq:varprinciple}
\delta \left[ \int d^{n}q\sqrt{g}\rho (g^{ij}\partial _{i}\sigma \partial
_{j}\sigma +R)\right] =0
\end{equation}%
where $g=|\det (g_{ij})|$, $R(q)$ is the scalar curvature and $\rho (q)$ and
$\sigma (q)$ are scalar fields. Variation with respect to $\rho (q)$ and $%
\sigma (q)$ yields, respectively~\cite{santamato85}
\begin{equation}
g^{ij}\partial _{i}\sigma \partial _{j}\sigma +R=0\hspace{1cm}(D_{k}\sigma
D^{k}\sigma +R=0)  \label{eq:HJ}
\end{equation}%
and
\begin{equation}
\frac{1}{\sqrt{g}}\partial _{i}(\sqrt{g}\rho g^{ij}\partial _{j}\sigma )=0%
\hspace{1cm}(D_{k}D^{k}\sigma =0).  \label{eq:cont}
\end{equation}%
Variation of (\ref{eq:varprinciple}) with respect to the connections $\Gamma
_{jk}^{i}(q)$ yields the Weyl conformal connection~\cite{weyl}
\begin{equation}
\Gamma _{jk}^{i}=-{i\brace jk}+\delta _{j}^{i}\phi _{k}+\delta _{k}^{i}\phi
_{j}+g_{jk}\phi ^{i},  \label{eq:Weylconnections}
\end{equation}%
where ${i\brace jk}$ are the Christoffel symbols out of the metric $g_{ij}$,
$\phi ^{i}=g^{ij}\phi _{j}$, and $\phi _{i}$ is Weyl's vector given by~\cite%
{santamato85}
\begin{equation}
\phi _{i}=-\frac{1}{n-2}\frac{\partial _{i}\rho }{\rho }\hspace{1cm}%
(D_{k}\rho =0).  \label{eq:rho}
\end{equation}%
The curvature tensor $R_{jkl}^{i}$ and the scalar curvature $R$ derived from
the connections (\ref{eq:Weylconnections}) are named the Weyl curvature
tensor and the Weyl scalar curvature, respectively. Moreover, Eq.~(\ref%
{eq:rho}) shows that the Weyl vector $\phi _{i}$ is a gradient, so that the
Weyl connection (\ref{eq:Weylconnections}) is integrable and we may take $%
\rho $ as Weyl's potential. Inserting Eq.~(\ref{eq:rho}) into the well-known
expression of Weyl's scalar curvature~\cite{weyl}, we obtain
\begin{equation}
R=\overline{R}+\left( \frac{n-1}{n-2}\right) \left[ \frac{g^{ij}\partial
_{i}\rho \partial _{j}\rho }{\rho ^{2}}-\frac{2\partial _{i}(\sqrt{g}%
g^{ij}\partial _{j}\rho )}{\rho \sqrt{g}}\right]  \label{eq:R}
\end{equation}%
where $\overline{R}$ is the Riemann curvature of $V_{n}$ calculated from the
Christoffel symbols of the metric $g_{ij}$. The connections~(\ref%
{eq:Weylconnections}) are invariant under the Weyl conformal gauge
transformations~\cite{weyl}
\begin{eqnarray}  \label{eq:Weylgauge}
g_{ij} &\rightarrow &\lambda g_{ij}  \label{eq:gtransform} \\
\phi _{i} &\rightarrow &\phi _{i}-\frac{\partial _{i}\lambda }{2\lambda }.
\label{eq:phitransform}
\end{eqnarray}%
The fields $T(q)$ which under Weyl-gauge transform as $T\rightarrow \lambda
^{w(T)}T$ are said to transform simply and the exponent $w(T)$ is the Weyl
"weight" of $T$. Examples are $w(g_{ij})=1$, $w(g^{ij})=-1$, $w(\sqrt{g}%
)=n/2 $ and $w(R)=-1$. the Weyl vector $\phi _{i}$ does not transform
simply, as shown by Eq.~(\ref{eq:phitransform}). We see that principle (\ref%
{eq:varprinciple}) is Weyl-gauge invariant provided $w(\sigma )=0$ and $%
w(\rho )=-(n-2)/2$. In the Weyl geometry is convenient to introduce the
Weyl's conformally-covariant (co-covariant) derivative $D_{i}$ so that the
metric tensor is constant, i.e. $D_{i}g_{jk}=0$. For a tensor field $T$ of
weight $w(T)$ we have $D_{i}T=\nabla _{i}^{(\Gamma )}T-2w(T)\phi _{i}T$,
where $\nabla _{i}^{(\Gamma )}$ is the covariant derivative derived from the
connections~(\ref{eq:Weylconnections}). The Weyl covariant derivative leaves
$w$ unchanged, i.e. $w(D_{i}T)=w(T)$. Because $D_{i}g_{jk}=0$, summation
indices can be raised and lowered using the metric, as usually made in the
Riemann geometry where the covariant derivative is $\nabla _{i}$. In the
parenthesis of Eqs.~(\ref{eq:HJ}),(\ref{eq:cont}),(\ref{eq:rho}) are the
same expressions in the co-covariant form so to make the Weyl-gauge
covariance of the theory explicit. We notice, in particular, that $\rho $ is
constant with respect to the co-covariant derivative. The field equations~(%
\ref{eq:HJ}), (\ref{eq:cont}), (\ref{eq:rho}), and (\ref{eq:R}) are the main
equations of the theory.
\section{The mechanical interpretation}
The field theory based on the variational principle (\ref{eq:varprinciple})
has a straightforward mechanical interpretation. In fact, the field Eq.~(\ref%
{eq:HJ}) has the form of the Hamilton-Jacobi Equation (HJE) of mechanics for
the action function $\sigma (q)$ of a particle subjected to the scalar
potential given by the Weyl curvature (\ref{eq:R}). Alternatively, we may
derive Eq.~(\ref{eq:HJ}) from the single-integral variational problem $%
\delta \int Ld\tau =0$ with the homogeneous Lagrangian~\footnote{%
We introduce here an homogeneous Lagrangian so to have a parameter invariant
action principle, as required by relativity.}
\begin{equation}
L(q,\dot{q})=\sqrt{-R(q)g_{ij}(q)\dot{q}^{i}\dot{q}^{j}}.  \label{eq:L}
\end{equation}%
This Lagrangian (and the associated HJE) have the same form of the
Lagrangian of a relativistic particle moving in space-time with mass
constant replaced by the curvature field $R(q)$. Any solution $\sigma (q)$
of the HJE defines a bundle of (time-like) trajectories in $V_{n}$ given by $%
\dot{q}^{i}=g^{ij}\partial _{j}\sigma $, corresponding to possible
trajectories of the system in the configuration space, when the system is in
the dynamical \textit{state} defined by $\sigma (q)$. Each trajectory of the
bundle obeys the Euler-Lagrange equations derived from $L$, so that along
its motion, the system is subjected to a newtonian force proportional to the
gradient of the Weyl curvature $R$. However, as said above, the dynamics
described by $\sigma (q)$ must be compatible with the affine connections of $%
V_{n}$ and, hence, the curvature potential $R$ as well as $\sigma $ must be
simultaneous solutions of Eqs.~(\ref{eq:HJ}) and (\ref{eq:cont}). Once these
two equation are solved, the field $\sigma (q)$ fixes the dynamics and the
field $\rho (q)$ fixes the affine connections from Eqs.~(\ref%
{eq:Weylconnections}) and (\ref{eq:rho}), and the curvature from Eq.~(\ref%
{eq:R}).\newline
In addition, the field equation (\ref{eq:cont}) has a simple mechanical
interpretation as a "continuity equation" $(\partial _{i}j^{i}=0)$ for the
current density
\begin{equation}
j^{i}=\sqrt{g}\:\rho g^{ij}\partial _{j}\sigma .  \label{eq:j}
\end{equation}%
It is worth noting that the current density $j^{i}$ has $w(j^{i})=0$ and is
therefore Weyl-gauge invariant (co-covariant). This is an important point in
a consistent conformally invariant approach, because it is expected that
only gauge-invariant quantities have definite physical meaning and can be
measured experimentally. We will return on the measurement issue in the
final part of the paper. Here we conclude by observing that the continuity
equation (\ref{eq:cont}) could also describe the motion of a fluid of
density $\rho $ conveyed along the bundle of trajectories defined by $\sigma
$ according to the hydrodynamical picture of quantum mechanics~\cite%
{madelung26}. Moreover, the last term on the right of Eq.~(\ref{eq:R}) has
the same mathematical form of the \textquotedblleft quantum
potential\textquotedblright introduced "ad hoc" by David Bohm in order to
derive the Schr\"{o}dinger's equation~\cite{bohm83a,bohm83b}. However, the
\textquotedblleft quantum potential\textquotedblright , whose gradient acts
as a newtonian force on the particle, has a quite mysterious origin: it was
in facts introduced "ad hoc" by Bohm. According to the present (CQG) theory,
the active potential originates from geometry, as does gravitation, and
arises from the space curvature due to the presence of the non trivial
affine connections of the Weyl's conformal geometry. Furthermore, it is also
worth noting that the conformal invariance requires that the Riemann scalar
curvature contributes to the potential: a contribution which is absent in
Bohm's approach.
\section{The scalar wavefunction}
We may exploit this formal analogy to simplify the nonlinear problem implied
by Eqs.~(\ref{eq:HJ}) and (\ref{eq:cont}) by introducing the complex field $%
\psi (q)$ given by
\begin{equation}
\psi (q)=\sqrt{\rho }e^{i\frac{S}{\hbar }}  \label{eq:ansatz}
\end{equation}%
with $S(q)=\xi \hbar \sigma (q)$, and
\begin{equation}
\xi =\sqrt{\frac{n-2}{4(n-1)}}.  \label{eq:xi}
\end{equation}%
With the \textit{ansatz} (\ref{eq:ansatz}) the field equations (\ref{eq:HJ})
and (\ref{eq:cont}) can be grouped in the single linear wave equation for
the complex field $\psi (q)$ given by
\begin{equation}
\Delta _{c}\psi \equiv (\nabla _{k}\nabla ^{k}-\xi ^{2}\overline{R})\psi =0,
\label{eq:wave}
\end{equation}%
where the $\Delta _{c}$ is the conformal Laplace operator, $\nabla
_{k}\nabla ^{k}$ is the Laplace-Beltrami operator and $\overline{R}$ is the
Riemann scalar curvature of $V_{n}$ calculated by the metric tensor $g_{ij}$%
. A striking circumstance follows from this approach. Namely, although Eq.~(%
\ref{eq:wave}) is mathematically equivalent to Eqs.~(\ref{eq:HJ}) and (\ref%
{eq:cont}), any direct reference to Weyl's geometric structure of $V_{n}$
formally disappears in a theory based on Eq.~(\ref{eq:wave}). This
remarkable feature, which affects all quantum equations obtained by CQG, i.e
Schr\"{o}dinger's or Dirac's, may explain why this or a similar theory based
on Weyl's results was never previously formulated. \ \ In facts, Eq.~(\ref%
{eq:wave})can be written directly once the metric tensor is known, without
any reference to the underlying affine connections (\ref{eq:Weylconnections}%
) and curvature (\ref{eq:R}). The form of Eq.~(\ref{eq:wave}) is the same in
all conformal gauges provided $w(\psi )=-(n-2)/4$ (or $\psi \rightarrow \psi
^{\prime }=\lambda ^{-(n-2)/4}\psi $), as it can be easily checked from the
well-known transformation law of the Riemann scalar curvature under the
conformal change $g_{ij}\rightarrow g_{ij}^{\prime }=\lambda g_{ij}$ of the
metric~~\cite{levicivita}. In other words, all information about Weyl's
structure of the configuration space is lost in the ensuing theory if the
wave equation (\ref{eq:wave}) is taken as the starting point of the theory:
a full knowledge of the dynamical features of the system may be gained only by
making recourse to the full set made by the two nonlinear Eqs.~(\ref{eq:HJ})
and (\ref{eq:cont}), and to the associated set of trajectories in $V_{n}$
subjected to the Weyl curvature potential. As it will be shown later, it is
precisely the Weyl potential which produces the quantum entanglement, thus
unveiling the true dynamical nature of the EPR "paradox". Therefore, the
wave function $\psi (q)$ expressed by Eq.~(\ref{eq:ansatz}) should be
considered as a no more than a useful mathematical \textit{ansatz }apt to
convert the fundamental set made by Eqs.~(\ref{eq:HJ}) and (\ref{eq:cont})
into a simpler linear "wave equation".
\section{Including external electromagnetic fields}
External electromagnetic fields are easily introduced in the theory by the
rule $\partial _{i}\sigma \rightarrow \partial _{i}\sigma -a_{i}$ applied to
Eqs.~(\ref{eq:HJ}), (\ref{eq:cont}) and (\ref{eq:j}), and by adding the term
$a_{i}(q)\dot{q}^i$ to the Lagrangian $L$ in Eq.~(\ref{eq:L}). In this way,
invariance is gained also with respect to the electromagnetic gauge changes $%
a_{i}\rightarrow a_{i}+\partial _{i}\chi $ and $\sigma \rightarrow \sigma
+\chi $. Finally, the Weyl conformal invariance requires $w(a_{i})=0$. When
the \textit{ansatz} (\ref{eq:ansatz}) is used, the wave equation (\ref%
{eq:wave}) is changed into
\begin{equation}
g^{ij}\left( \hat{p}_{i}-\frac{e}{c}A_{i}\right) \left( \hat{p}_{j}-\frac{e}{%
c}A_{j}\right) \psi +\hbar ^{2}\xi ^{2}\overline{R}\psi =0  \label{eq:waveem}
\end{equation}%
where we may set $a_{i}=\frac{e}{\hbar \xi c}A_{i}$, $\hat{p}_{k}=-i\hbar
\nabla _{k}$ in order to obtain a more familiar appearance of the wave
equation as an $n$-dimensional Klein-Gordon equation with the mass term
replaced by the Riemann scalar curvature of $V_{n}$. With the same
notations, the dynamical Eqs.~(\ref{eq:HJ}) and (\ref{eq:cont}) become
\begin{eqnarray}
g^{ij}\left( \partial _{i}S-\frac{e}{c}A_{i}\right) \left( \partial _{j}S-%
\frac{e}{c}A_{j}\right) +\hbar ^{2}\xi ^{2}R &=&0  \label{eq:HJem} \\
\frac{1}{\sqrt{g}}\partial _{i}\left[ \sqrt{g}\rho g^{ij}\left( \partial
_{j}S-\frac{e}{c}A_{j}\right) \right] &=&0.  \label{eq:contem}
\end{eqnarray}%
where all "quantum effects" are accounted for by the Weyl curvature term in
Eq.~(\ref{eq:HJem}), which vanishes in the "classical" limit: $\hbar
\rightarrow 0$.\\
\section{The relativistic spinning particle}
Spin is one of the cornerstones of quantum mechanics. Consequently, being
the spin a peculiar feature of the quantum world, any attempt to find a
classical system behaving as a spinning quantum particle is generally
considered hopeless. Equations (\ref{eq:HJem}) and (\ref{eq:contem}) have a
classical structure and the wave equation (\ref{eq:waveem}) has only the
look of a "quantum equation". Since the last equation has the typical
\textquotedblleft bosonic\textquotedblright form, it is not very surprising
that Eqs.~ (\ref{eq:HJem}), (\ref{eq:contem}) and (\ref{eq:waveem}) may
reproduce all details of the behavior of a quantum integer spin. However, it
may be indeed surprising that even the half-integer spin may be accounted
for by (\ref{eq:waveem}). The proof of this statement is the subject of the
present section.\newline

We start from the model by Boop and Haag~\cite{bopp50} of a relativistic top
described by six Euler angles $\theta ^{A}$ $(A=1,\dots ,6)$. We may
visualize this top as a rigid fourleg $e_{a}^{\mu }(\theta )$ $(\mu
,a=0,\dots ,3) $ parametrized by the six angles $\theta ^{A}$ whose origin
is located at point $x^{\mu }$ in Minkowski space-time with metric tensor $%
g_{\mu\nu }=\mathrm{diag}(-1,1,1,1)$. The fourleg vectors $e_{a}^{\mu
}(\theta )$ are normalized so that $g^{\mu \nu }e_{a}^{\mu }e_{b}^{\nu
}=\gamma _{ab}=\mathrm{diag}(-1,1,1,1)$. With some abuse of language, we may
say that the coordinates $x^{\mu }$ belong to the center of mass of the top
and that the angles $\theta ^{A}$ yield the top \textquotedblleft
orientation\textquotedblright in space-time, even if the vector $e_{0}^{\mu
} $ of the fourleg is time-like. We assume also the time component $%
e_{0}^{0} $ of $e_{0}^{\mu }$ positive, so that the matrix $\Lambda
=\{e_{a}^{\mu }\}$ is an orthochronous proper Lorentz matrix. The motion of
the fourleg is described by the world line $x^{\mu }(\tau )$ of its center
of mass and by the motion of the four vectors $e_{a}^{\mu }(\tau )$
described by the six functions $\theta ^{A}(\tau )$. The parameter $\tau $
is arbitrary, but sometimes it is convenient to take as parameter the
space-time arc element $ds$ given by $-ds^{2}=g_{\mu \nu }dx^{\mu }dx^{\nu }$%
. The four-velocity of the center of mass is given by $u^{\mu }=dx^{\mu }/ds$
and the \textquotedblleft angular velocity\textquotedblright of the fourleg $%
e_{a}^{\mu }$ is given by the tensor $\omega _{\nu }^{\mu }$ defined by $%
de_{a}^{\mu }/ds=\omega _{\nu }^{\mu }e_{a}^{\nu }$. If the parameter $\tau$ is
chosen gauge invariant, we have $w(\omega^{\mu}_{\nu})=0$. From normalization we
obtain $u_{\mu }u^{\mu }=-1$ and $\omega ^{\mu \nu }+\omega ^{\nu \mu }=0$,
i.e. $\omega ^{\mu \nu }=g^{\rho \nu }\omega _{\rho }^{\mu }=g^{\rho \nu
}e_{\rho }^{a}de_{a}^{\mu }/d\tau $ is antisymmetric ($e_{\mu }^{a}$ are the
reciprocal elements of $e_{a}^{\mu }$, i.e. $e_{\mu }^{a}e_{b}^{\mu }=\delta
_{b}^{a}$). The configuration space of the relativistic top is the
ten-dimensional space $V_{10}$ and Eq.~(\ref{eq:xi}) yields $\xi =\sqrt{2}/3$%
. The metric tensor $g_{ij}$ of $V_{10}$ has a two-block diagonal form. In
the first upper block is the Minkowski metric $g^{\mu \nu }=\mathrm{diag}%
(-1,1,1,1)$ and the last lower block is given by the $6\times 6$ Euler angle
metric tensor $\gamma _{AB}(\theta )=-a^{2}g_{\mu \rho }g_{\nu \sigma
}\omega _{A}^{\mu \nu }(\theta )\omega _{B}^{\rho \sigma }(\theta )$ where $%
\omega _{A}^{\mu \nu }(\theta )=g^{\rho \nu }e_{\rho }^{a}(\theta )\partial
_{\theta ^{A}}e_{a}^{\mu }(\theta )$.\newline
According to the general principles of CQG, we generalize the Lagrangian
introduced by Bopp and Haag~\cite{bopp50} to:
\begin{equation}
L=\sqrt{-\hbar ^{2}\xi ^{2}R(g_{\mu \nu }\dot{x}^{\mu }\dot{x}^{\nu
}-a^{2}\omega _{\mu \nu }\omega ^{\mu \nu })}+\frac{e}{c}A_{\mu }\dot{x}%
^{\mu }+\frac{\kappa a^{2}e}{c}F_{\mu \nu }\omega ^{\mu \nu },
\label{eq:Ltop}
\end{equation}%
where a gauge ivariant parameter $\tau$ is assumed, $R$ is Weyl's curvature of
$V_{10}$, $a$ is the top \textquotedblleft gyration
radius\textquotedblright\ with $w(a^2)=1$, $e$ is the top charge, $%
A_{\mu }$ is the electromagnetic four potential, $F_{\mu \nu }=\partial
_{\mu }A_{\nu }-\partial _{\nu }A_{\mu }$ is the electromagnetic tensor and,
finally, $\kappa $ is a numerical coupling constant~\cite%
{santamato13,demartini11}. When written in full as a function of the ten
generalized coordinates $q^{i}=\{x^{\mu },\theta ^{A}\}$ and their
derivatives, the Lagrangian (\ref{eq:Ltop}) reduces to the canonical form~(%
\ref{eq:L}) with the addition of the electromagnetic term $a_{i}(q)\dot{q}%
^{i}$ and vector $a_{i}(q)=\{a_{\mu }(x),a_{A}(x,\theta )\}=(\hbar \xi
)^{-1}\{\frac{e}{c}A_{\mu },\frac{\kappa a^{2}e}{c}F_{\mu \nu }\omega
_{A}^{\mu \nu }\}$. Therefore, the dynamical equations~(\ref{eq:HJem}) and (%
\ref{eq:contem}), the \textit{ansatz} (\ref{eq:ansatz}), and the wave
equation (\ref{eq:waveem}) apply. Unlike Minkowski space-time, which is
flat, the configuration space $V_{10}$ is curved and has a constant Riemann
curvature $\overline{R}=6/a^{2}$. We see, therefore, that a constant mass
appears in the wave equation (\ref{eq:waveem}) of the spinning particle.
However, Eq.~(\ref{eq:waveem}) still has its \textquotedblleft
bosonic\textquotedblright character. To gain a connection with the spinorial
description adopted in traditional quantum mechanics, we seek for solutions $%
\psi (q)$ of Eq.~(\ref{eq:waveem}) in the mode expansion form
\begin{equation}
\psi _{uv}(q)=D^{(u,v)}(\Lambda ^{-1}(\theta ))_{\sigma }\psi ^{\sigma
}(x)+D^{(v,u)}(\Lambda ^{-1}(\theta ))_{\dot{\sigma}}\psi ^{\dot{\sigma}}(x)%
\hspace{0.8cm}(u\leq v)  \label{eq:psiuv}
\end{equation}%
where $D^{(u,v)}(\Lambda (\theta ))_{\sigma }$ is the first raw of the $%
(2u+1){\times }(2v+1)$ matrix representing the Lorentz transformation $%
\Lambda (\theta )=\{e_{a}^{\mu }(\theta )\}$ in the irreducible
representation labeled by the two numbers $u,v$ given by $2u,2v=0,1,2,\dots $%
, and the $\psi ^{\sigma }(x)$ and $\psi ^{\dot{\sigma}}(x)$ are expansion
coefficients depending on the space-time coordinates $x^{\mu }$ only. The
matrices $D^{(u,v)}(\Lambda (\theta ))$ and $D^{(v,u)}(\Lambda (\theta ))$
depend on the Euler angles $\theta ^{A}$ only, and provide conjugate
representations of the Lorentz transformations~\footnote{%
The two matrices are related by $[D^{(u,v)}(\Lambda )]^{\dag
}=[D^{(v,u)}(\Lambda )]^{-1}$.}. As suggested by the notation, the
invariance of $\psi _{uv}(q)$ under Lorentz transformations implies that $%
\psi ^{\sigma }(x)$ and $\psi ^{\dot{\sigma}}(x)$ change as undotted and
dotted contravariant spinors, respectively~\footnote{%
The spinors $\psi ^{\sigma }(x)$ and $\psi ^{\dot{\sigma}}(x)$ have a second
Lorentz invariant lower index $\sigma ^{\prime }$ and $\dot{\sigma}%
^{\prime }$, respectively, related to the spin component $s_{\zeta }$ along
the top moving axis $\zeta $. With no loss of generality we may orient the
axis so to have $s_{\zeta }$ fixed and omit $\sigma ^{\prime }$ and $\dot{%
\sigma}^{\prime }$}. Insertion of the expansion (\ref{eq:psiuv}) into the
wave-equation (\ref{eq:waveem}) yields the following equation for the
coefficients $\psi ^{\sigma }(x)$ and $\psi ^{\dot{\sigma}}(x)$
\begin{equation}
\left[ g^{\mu \nu }\left( \hat{p}_{\mu }-\frac{e}{c}A_{\mu }\right) \left(
\hat{p}_{\nu }-\frac{e}{c}A_{\nu }\right) +\hbar ^{2}\xi ^{2}\overline{R}%
\right] \psi (x)+\Delta _{J}\psi (x)=0  \label{eq:coeff}
\end{equation}%
where $\overline{R}=6/a^{2}$, $\psi (x)$ denotes either $\psi ^{\sigma }(x)$
or $\psi ^{\dot{\sigma}}(x)$ and $\Delta _{J}$ is a $(2u+1){\times }(2v+1)$
matrix depending on the space-time coordinates $x^{\mu }$ only, given by
\begin{equation}
\Delta _{J}=\left[ \frac{\hbar }{a}\bm J-\frac{\kappa ea}{2c}\bm H\right]
^{2}-\left[ \frac{\hbar }{a}\bm K-\frac{\kappa ea}{2c}\bm E\right] ^{2}.
\label{eq:DeltaJ}
\end{equation}%
Here $\bm J$ and $\bm K$ are the generators of the Lorentz group in the
undotted (or dotted) conjugate representation, corresponding to $\psi
^{\sigma }(x)$ (or to $\psi ^{\dot{\sigma}}(x)$). We notice that the motion
of the rotating fourleg described by the HJE~(\ref{eq:HJem}) is in the group
SO(3,1) of proper Lorentz transformations, while the evolution of the
spinors $\psi ^{\sigma }(x)$ and $\psi ^{\dot{\sigma}}(x)$ is in the group
of complex $D$-matrices. This last motion, however, has only an auxiliary
role in the present approach, where the physics is ascribed to the fourleg
dynamics.\\

Before concluding this section, we observe that the choice of the Minkowski
space-time metric $g_{\mu\nu }=\mathrm{diag}(-1,1,1,1)$ can be made only in
one gauge, that we can call the "Minkowski gauge". Only in this gauge the
comparison of CQG and standard Quantum Mechanics can be made
and only in this gauge the quantum effects and gravitational effects can be
ascribed to independent geometric concepts: vector parallel transport and
vector length, respectively. In other gauges, a clear separation is
impossible and a different picture may emerge. This happens, for example,
in the gauge used originally by Weyl in its approach to electromagnetism,
where Weyl's curvature $R$ is constant (we may call this gauge the "Weyl gauge").
It is however worth noting that a also a gauge exists where both quantum
and gravitational phenomena share the same origin in the metric only. In
this particular gauge, that we can call the "Riemann gauge" the geometry is
pure Riemann and is entirely governed by the metric tensor~\footnote{%
The existence of the Riemann gauge is due to the fact that in our case
the Weyl connections are integrable}. In the Riemann
gauge, we have $\rho=\mathrm{const.}$, the Weyl vector vanishes and the
Weyl curvature reduces to the Riemann curvature built from the the metric
$\bar g_{ij}$ given by $\bar g_{ij}=|\Psi(q)|^{\frac{4}{n-2}}g_{ij}$.
Notice, however,that the space-time upper diagonal block of the metric
$\bar g_{ij}$ depends now on the space-time coordinates and on the Euler
angles as well.
\section{The relativistic spin $\frac{1}{2}$}
Equation (\ref{eq:coeff}) is written for \textit{any} spin. Spin $\frac12$
is obtained by setting $u=0$ and $v=\frac12$ in Eq.~(\ref{eq:psiuv}) so that
$D^{(0,\frac12)}(\Lambda (\theta ))$ and $D^{(\frac12,0)}(\Lambda (\theta
))\in SL(2,C)$ and $\psi ^{\sigma }(x)$ and $\psi ^{\dot{\sigma}}(x)$ are
two component undotted and dotted Lorentz spinors, respectively. Then,
introducing the Dirac four component spinors $\Psi _{D}={\psi ^{\sigma }%
\brace\psi ^{\dot{\sigma}}}$ and $\Phi _{D}={D(\theta )^{\sigma }\brace %
D(\theta )^{\dot{\sigma}}}$, where $D(\theta )^{\sigma }$ and $D(\theta )^{%
\dot{\sigma}}$ are the first column of the matrices $D^{(0,\frac12)}(\Lambda
(\theta ))$ and $D^{(\frac12,0)}(\Lambda (\theta ))$, respectively, Eq.~(\ref%
{eq:psiuv}) can be written as the Dirac product $\psi (q)=\overline{\Phi }%
_{D}(\theta )\Psi _{D}(x)=\Phi _{D}^{\dag }(\theta )\gamma ^{0}\Psi _{D}(x)$%
, where $\gamma ^{0}={0\;1\brace1\;0}$ is Dirac's matrix in the spinor
representation. Moreover, setting $\kappa =2$ for the electron, Eq.~(\ref%
{eq:coeff}) yields:
\begin{eqnarray}
\left[ g^{\mu \nu }\left( \hat{p}_{\mu }-\frac{e}{c}A_{\mu }\right) \left(
\hat{p}_{\nu }-\frac{e}{c}A_{\nu }\right) -\frac{e\hbar }{c}(\bm\Sigma {%
\cdot }\bm H-i\bm\alpha {\cdot }\bm E)+\frac{3\hbar ^{2}}{2a^{2}}(1+4\xi
^{2})\right] \Psi _{D} +  \nonumber \label{eq:Dirac} \\
\mbox{}+\left[ \frac{e^{2}a^{2}}{c^{2}}(H^{2}-E^{2})\right] \Psi _{D}=0,%
\hspace{3cm} &&
\end{eqnarray}%
where $\bm\Sigma ={\bm\sigma \;0\brace0\;\bm\sigma }$, $\bm\alpha ={\bm%
\sigma \;\;0\brace0\;-\bm\sigma }$, and $\bm\sigma =\{\sigma _{x},\sigma
_{y},\sigma _{z}\}$ are the usual Pauli matrices. By setting
\begin{equation}
a=(\hbar /mc)\sqrt{3(1+4\xi ^{2})/2},  \label{eq:a}
\end{equation}%
where $m$ is the electron mass, and by neglecting the term $%
(ea/c)^{2}(H^{2}-E^{2})$, Eq.~(\ref{eq:Dirac}) reduces to the second-order
(squared) Dirac's equation in the spinor representation~[see, for example,
Ref.~\cite{landau4}, Eq.~(32,7a)]. A more compact form of Eq.~(\ref{eq:Dirac}%
) is~\cite{landau4}
\begin{equation}
\left[ \gamma ^{\mu }\gamma ^{\nu }\left( \hat{p}_{\mu }-\frac{e}{c}A_{\mu
}\right) \left( \hat{p}_{\nu }-\frac{e}{c}A_{\nu }\right) -m^{2}c^{2}\right]
\Psi _{D}=0,  \label{eq:Diracgamma}
\end{equation}%
where $\gamma ^{\mu }$ are Dirac's matrices in the spinor representation. As
it is well known, Eq.~(\ref{eq:Diracgamma}) can be written as $\hat{\mathcal{%
D}}_{+}\hat{\mathcal{D}}_{-}\psi _{D}=\hat{\mathcal{D}}_{-}\hat{\mathcal{D}}%
_{+}\psi _{D}=0$, where $\hat{\mathcal{D}}_{\pm }=\gamma ^{\mu }(p_{\mu
}-(e/c)A_{\mu })\pm m$ are first-order Dirac's operators with positive and
negative mass $m$, respectively. Any solution $\Psi _{D}$ of the
second-order Eq.~(\ref{eq:Diracgamma}) can be written as a linear
superposition of a solution $\Psi _{+}$ of the first order Dirac's equation $%
\hat{\mathcal{D}}_{+}\Psi _{+}=0$ with positive mass $m$ and a solution of
the first-order equation $\hat{\mathcal{D}}_{-}\Psi _{-}=0$ with negative
mass. To have full correspondence with the first-order Dirac's equation,
negative mass solutions of Eq.~(\ref{eq:Diracgamma}) must be disregarded as
unphysical because they correspond to particles affected by an improper
boost (negative determinant) from rest-frame. A systematic way to drop out
the unphysical negative mass solutions is to start from arbitrary four
component solution $\Psi _{D}$ of the second-order equation (\ref%
{eq:Diracgamma}) and define the field $\Phi _{D}=\hat{\mathcal{D}}_{-}\Psi
_{D}$. Then $\Phi _{D}$, besides being a solution of Eq.~(\ref{eq:Diracgamma}%
) is also a solution of the first-order Dirac's equation~[see Ref.~\cite%
{landau4}, Sec. 32]. The occurrence of second order Dirac's equation (\ref%
{eq:Diracgamma}) is expected in the present approach because of the
\textquotedblleft bosonic\textquotedblright character of Eq.~(\ref{eq:waveem}%
). We introduced here four component Dirac spinors because we required
invariance under parity transformation. However, it is worth noting that the
wave equation~(\ref{eq:waveem}) has also chiral solutions. In fact each one
of the two terms on the right of Eq.~(\ref{eq:psiuv}) obeys Eq.~(\ref%
{eq:waveem}). These solutions correspond to two-component Lorentz spinors
with opposite chirality and may have a role in no-parity-preserving
interactions. Moreover, as shown by Brown~\cite{brown58}, the two-component
solutions, beside reproducing the same physical results of Dirac's equation
when parity is restored, are also computationally easier to work with.
Finally, we notice the presence of the last term on the right of Eq.~(\ref%
{eq:Dirac}), which is absent in the standard second order Dirac equation~(%
\ref{eq:Diracgamma}). This term quadratic in the applied fields is needed to
preserve the Weyl conformal invariance of the underlying theory and cannot
be suppressed. However, the contribution of this term in the equation is
negligibly small. In fact, Eq.~(\ref{eq:a}) shows that $a$ is of the order
of the electron Compton wavelength $\lambda _{C}$. We may then estimate the
field $E$ required to render the quadratic term in Eq.~(\ref{eq:Dirac})
comparable with the linear one. We find: $E\simeq 10^{18}$~V/m. To have an
idea how large is this field, an electron at rest is accelerated by such
field up to $10^{9}$~GeV in a linear accelerator 1~m long. Similarly the
term quadratic in the magnetic field becomes comparable with the linear one
for the extremely large field: $H\simeq 10^{9}$~T.
\section{The nonrelativistic limit}
As we have seen, any positive mass solution of the second order Dirac
equation (\ref{eq:Dirac}) provides the coefficients $\psi ^{\sigma }(x)$ and
$\psi ^{\dot{\sigma}}(x)$ in the mode expansion (\ref{eq:psiuv}) of the
wavefunction $\psi (q)$. Taking modulus and phase of $\psi (q)$ we can find
the corresponding solution of our main Eqs.~(\ref{eq:HJem}) and (\ref%
{eq:contem}) which fix the dynamics of the system and the compatible Weyl
geometry of the configuration space. The HJE (\ref{eq:HJem}), in particular,
defines a bundle of paths $\{x^{\mu }(\tau ),e_{a}^{\mu }(\tau )\}$ in the
configuration space. The curves $x^{\mu }(\tau )$ correspond to the world
lines described by the \textquotedblleft center of mass\textquotedblright\
of the particle with four-velocity $u^{\mu }=\{u^{0},\bm u\}=\frac{dx^{\mu }%
}{ds}(\tau )$. The motion of the fourleg $e_{a}^{\mu }(\tau )$ defines a
rotation of the three space-like unit vectors $\{e_{1}^{\mu },e_{2}^{\mu
},e_{3}^{\mu }\}$ along the orthogonal axes $\xi ,\eta ,\zeta $ co-moving
with the particle, while the time-like vector $e_{0}^{\mu }(\tau )$
describes the world line $y^{\mu }(\tau )$ of the particle \textquotedblleft
center of energy\textquotedblright with four-velocity given by $v^{\mu }=%
\frac{dy^{\mu }}{ds}(\tau )=e_{0}^{\mu }(\tau )$. In general, $u^{\mu }$ and
$v^{\mu }$ are different, a phenomenon known as \textit{zitterbewegung}. The
dynamics of such classical rotating object described by six Euler angles can
be found, e.g. in the book by Sudarshan and Mukunda~\cite{sudarshan},
Chap.~20. However, the detailed study of this motion and of the \textit{%
zitterbewegung} is beyond the scope of the present work and will be left for
future work. Here we limit to study the nonrelativistic limit of the theory
when velocities are much lower than the speed of light. To this purpose, it
is convenient to factorize the Lorentz transformation $\Lambda (\theta
)=\{e_{a}^{\mu }(\theta )\}$ associated to the particle fourleg as $\Lambda
(\theta )=B(e_{0})R(\alpha ,\beta ,\gamma )$ where $R(\alpha ,\beta ,\gamma
) $ is a rotation matrix $\in SO(3)$ depending in the three Euler angles $%
\{\alpha ,\beta ,\gamma \}$ and $B(e_{0})$ is the boost associated to the
time-like vector $e_{0}^{\mu }$ of the particle fourleg. The rotation $%
R(\alpha ,\beta ,\gamma )$ belongs to the little Poincar\'{e} group around $%
e_{0}^{\mu }$ and in a Lorentz transformation $\bar{\Lambda}$ the angles $%
\{\alpha ,\beta ,\gamma \}$ transform according to the Wigner rotation $%
B^{-1}(\bar{e}_{0})\bar{\Lambda}_{0}B(e_{0})$, where $\bar{e}_{0}^{\mu }=%
\bar{\Lambda}_{\rho }^{\mu }e_{0}^{\rho }$. When the factorization $\Lambda
(\theta )=B(e_{0})R(\alpha ,\beta ,\gamma )$ is inserted into Eq.~(\ref%
{eq:psiuv}) and spin $\frac12$ is considered, we obtain
\begin{eqnarray}
\psi (q) &=&[D(R^{-1}(\alpha ,\beta ,\gamma ))D(B^{-1}(e_{0}))]_{\sigma
}\psi ^{\sigma }(\bm r,t)+  \nonumber  \label{eq:psispin} \\
&&\mbox{}+[D(R^{-1}(\alpha ,\beta ,\gamma ))D(B(e_{0}))]_{\dot{\sigma}}\psi
^{\dot{\sigma}}(\bm r,t)
\end{eqnarray}%
where $D(R(\alpha ,\beta ,\gamma ))\in $~SU(2) and the boost $D(B(e^{0}))$
is given by $D^{2}(B(e^{0}))=e_{0}^{\mu }\sigma _{\mu }$ with $\sigma _{\mu
}=\{1,\bm\sigma \}$ the four-vector of Pauli's matrices. The non
relativistic limit is obtained from Eq.~(\ref{eq:psispin}) by setting $\psi
^{\dot{\sigma}}(x)\simeq \psi ^{\sigma }(x)=w^{\sigma }(\bm r,t)$, where $%
w^{\sigma }(\bm r,t)$ is a rotation two-component spinor, and setting $%
e_{0}^{\mu }=\{e_{0}^{0},\bm e\}\simeq (1,0,0,0)$ because the center of mass
velocity $v_{c}=\frac{c|\bm u|}{u_{0}^{0}}\ll c$ and the center of energy
velocity $v_{e}=\frac{c|\bm e|}{e_{0}^{0}}\ll c$. Then, the non relativistic
limit of Eq.~(\ref{eq:psispin}) is:
\begin{eqnarray}
\psi (q) &=&D(\alpha ,\beta ,\gamma )_{\sigma }w^{\sigma }(\bm %
r,t)=D_{\uparrow }(\alpha ,\beta ,\gamma )w_{\uparrow }(\bm %
r,t)+D_{\downarrow }(\alpha ,\beta ,\gamma )w_{\downarrow }(\bm r,t)=  \nonumber
\label{eq:psisingle} \\
&=&e^{i\frac{\gamma }{2}}\left[ e^{i\frac{\alpha }{2}}\cos \frac{\beta }{2}%
w_{\uparrow }(\bm r,t)+e^{-i\frac{\alpha }{2}}\sin \frac{\beta }{2}%
w_{\downarrow }(\bm r,t)\right],
\end{eqnarray}%
where, for brevity, we posed $D(\alpha ,\beta ,\gamma )=D(R^{-1}(\alpha
,\beta ,\gamma ))$~\footnote{%
An unessential phase factor $e^{-i\Omega t}$ with $\Omega =21\hbar
/(40ma^{2})$ should be inserted in Eq-~(\ref{eq:psisingle}) so tho make $%
w_{\uparrow }(\bm r,t)$ and $w_{\downarrow }(\bm r,t)$ obey Schr\"{o}dinger
equation. This phase factor will be omitted everywhere henceforth.}. At the
same time, Eq.~(\ref{eq:waveem}) reduces to the non relativistic Schr\"{o}%
dinger equation for Pauli's two-component spinor $w^{\sigma }(\bm %
r,t)=\{w_{\uparrow }(\bm r,t),w_{\downarrow }(\bm r,t)\}$ with components
corresponding to spin up or down along the fixed $z$-axis, respectively. The
configuration space is then reduced to the space $V_{6}$ spanned by the
position coordinates $\bm r$ and Euler angles $\{\zeta^a\}=\{\alpha ,\beta
,\gamma \}$ $(a=1,2,3)$. To better see the role played by the wavefunction
in the present approach, we consider the simple case of spin-up state. From
Eq.~(\ref{eq:psisingle}) with $w_\downarrow =0$ we calculate the mechanical
action $S$ and the Weyl curvature $R$ when the spin is up:
\begin{eqnarray}
S(\bm r,t,\zeta) &=&\frac{\hbar }{2}(\gamma +\alpha )+\arg (w_{\uparrow }(\bm r,t))
\label{eq:Ssingleup} \\
R(\bm r,t,\zeta) &=&-\frac{5}{2a^{2}(1+\cos \beta )}+R_{\uparrow }(\bm r,t)+\mathrm{const.}
\label{eq:Rsingleup}
\end{eqnarray}%
where $R_{\uparrow }(\bm r,t)$ is the contribution of $w_{\uparrow }(\bm %
r,t) $ to Weyl's curvature. From Eq.~(\ref{eq:Ssingleup}) we see that the $%
\beta $ coordinate is cyclic, and, hence, $\beta $ ia a constant of motion.
From Eq.~(\ref{eq:Rsingleup}) we see that the particle is \textit{not} free,
but is subjected to a self-force proportional to the gradient of the Weyl's
curvature. This self-force has a geometric origin and cannot be eliminated
since it is needed to have Weyl's gauge-invariance. However, its existence
is hidden in the standard quantum mechanics based on the space-time spinor $%
w_{\uparrow }(\bm r,t)$, which obeys the Schr\"{o}dinger equation for the
free particle. Similar considerations can be done for the spin-down state.
The non relativistic limit is much simpler to handle, so we will use Eq.~(%
\ref{eq:psisingle}) to investigate the intriguing problem raised by
Einstein, Podolsky and Rosen (EPR) in 1935, i.e. the famous, striking
phenomenon of "quantum nonlocality"~\cite{EPR35}.
\section{The two identical spin $\frac{1}{2}$ particles}
Following the EPR approach~\cite{EPR35}, we consider here two identical spin
$\frac{1}{2}$ nonrelativistic particles in the absence of external fields in
the nonrelativistic limit. The calculation to obtain Eqs.~(\ref{eq:Ssingleup}%
) and (\ref{eq:Rsingleup}) from the wavefunction (\ref{eq:psisingle}) can be
repeated when two identical spin $\frac{1}{2}$ particle are considered. The
configuration space is now the product space spanned by the 12 coordinates
given by the 6 space coordinates and the 6 angular coordinates of the two
particles. To clarify the source of EPR quantum correlations, we consider
here two cases: a) the two particles have opposite spin along the $z$-axis;
b) the two particles are in the EPR state. In the quantum notation, case (a)
correspond to the spin product state $|\uparrow \rangle |\downarrow \rangle $
and case (b) to the entangled state $(1/\sqrt{2})(|\uparrow \rangle
|\downarrow \rangle -|\downarrow \rangle |\uparrow \rangle )$.
\section{The product state of two opposite spins}
The wavefunction of the state $|\uparrow \rangle |\downarrow \rangle $ is
easily written by taking the product of the two terms on the right of Eq.~(%
\ref{eq:psisingle}) and $S$ and $R$ are then calculated from modulus and
phase of this wavefunction (for details see Ref.~(~\cite{santamato85}). The
result is
\begin{eqnarray}  \label{eq:updown}
\psi _{\uparrow \downarrow }(q) &=&D_{\uparrow }(\alpha _{A},\beta
_{A},\gamma _{A})D_{\downarrow }(\alpha _{B},\beta _{B},\gamma
_{B})w_{\uparrow }(\bm r_{A},t)w_{\downarrow }(\bm r_{B},t)
\label{eq:psiupdown} \\
S &=&S^{(A)}(\bm r_{A},t,\zeta _{A})+S^{(B)}(\bm r_{B},t,\zeta_{B})
\label{eq:Supdown} \\
R &=&R^{(A)}(\bm r_{A},t,\zeta _{A})+R^{(B)}(\bm r_{B},t,\zeta_{B})
\label{eq:Rupdown}
\end{eqnarray}%
where $S^{(A,B)}(\bm r_{A,B},t,\zeta _{A,B})$ and $R^{(A,B)}(\bm %
r_{A,B},t,\zeta _{A,B})$ are given respectively by Eqs.~(\ref{eq:Ssingleup})
and (\ref{eq:Rsingleup}) calculated for particle $A$ and $B$ separately.
From Eqs.~(\ref{eq:updown}) we se that in this case the particles have
independent motions. In particular, the Weyl curvature reduces to the sum of
the two Weyl curvatures so that each particle is affected only by its own
geometric self-force.
\section{The entangled two spin EPR state}
The same procedure can be applied to the EPR singlet wavefunction of the two
spins given by
\begin{equation}
\psi _{AB}(q)=\frac{1}{\sqrt{2}}(\psi _{\uparrow \downarrow }(q)-\psi
_{\downarrow \uparrow }(q))  \label{eq:psiEPR}
\end{equation}%
where $\psi _{\uparrow \downarrow }(q)$ is given by Eq.~(\ref{eq:updown})
and $\psi _{\downarrow \uparrow }(q)$ is obtained from this by exchanging
the up and down arrows. The result is~\cite{santamato85}
\begin{eqnarray}
S &=&\hbar \left[ \frac{\gamma _{A}+\gamma _{B}}{2}+\arctan \left( \csc
\frac{\beta _{A}-\beta _{B}}{2}\sin \frac{\beta _{A}+\beta _{B}}{2}\tan
\frac{\alpha _{B}-\alpha _{A}}{2}\right) \right. +  \nonumber  \label{eq:SEPR}
\\
&&\left. \mbox{}+\arg (w_{\uparrow }^{(A)}(\bm r_{A},t))+\arg (w_{\downarrow
}^{(B)}(\bm r_{B},t))\right]
\end{eqnarray}%
and
\begin{eqnarray}
R &=&\frac{22}{5a^{2}(1-\cos \beta _{A}\cos \beta _{B}-\cos \Delta \alpha
\sin \beta _{A}\sin \beta _{B})}+  \nonumber  \label{eq:REPR} \\
&&+R^{(A)}(\bm r_{A},t)+R^{(B)}(\bm r_{B},t)
\end{eqnarray}%
In this case, although the particle motions over the spatial "\textit{%
external variables}"$\left\{ x^{i}\right\} $ are independent, the particles
are still coupled by the Weyl curvature through the angular "\textit{%
internal variables}", $\left\{ \zeta^a_A,\zeta^b_B\right\}$ $(a,b=1,2,3)$
to the self-force, each one of them exerts a force on the other. We
conjecture, from our present limited nonrelativisic standpoint, that the
space-time superluminality of the nonlocal correlations comes from the
geometrical independency, i.e. disconnectedness, of the two $\left\{
x^{i}\right\} $ and $\left\{ \zeta^a_A,\zeta^b_B\right\}$ manifolds. The
superluminality issue indeed requires a fully relativistic future analysis.%
\newline

In the next two sections, we will consider in detail the behavior of the two
particles prepared in the EPR state (\ref{eq:psiEPR}), analyzed by a couple
of equal Stern-Gerlach Apparata (SGA). We will show that this geometrical
interaction among Euler's angles reproduces exactly all results of standard
quantum mechanics leading, in particular, to the violation of Bell's
inequalities.
\section{The meaning of the quantum measurement}
Any experimental apparatus designed to measure some physical property of a
quantum particle is made of two parts: 1) a \textquotedblleft
filtering"\textquotedblright device which addresses the particle to the
appropriate detector channel according the possible values of the quantity
to be measured (a spin component, in our case), 2) one (or more) detectors
able to register the arrival of the particle over each channel. To fix the
ideas, we consider here the particular case of the measure of a spin $%
\frac12 $ particle by a (SGA) apparatus. The spin component along the SGA
axis can have two values, so we need two detectors $D_{u}$ and $D_{d}$
coupled to the \textquotedblleft up\textquotedblright and \textquotedblleft
down\textquotedblright output channels of the orientable SGA. Each detector
measures the flux $\Phi $ of particles entering its acceptance area $A$.
Let's assume single particle detection. Then this flux is given by
\begin{equation}
\Phi =\textstyle\int_{\Sigma }j^{i}n_{i}d\Sigma =\textstyle\int_{\Sigma
}\rho \sqrt{g}g^{ij}\partial _{j}Sn_{i}d\Sigma  \label{eq:fluxVn}
\end{equation}%
extended to the hypersurface $\Sigma $ in the particle configuration space $%
V_{6}$ with normal unit vector $n_{i}=\{\bm{n},0,0,0\}$ where $\bm{n}$ is
the usual 3D-normal to the detector area $A$. Let us assume that the scalar
wavefunction of the particle at the detector location has its spacetime and
angular parts factorized, i.e. $\psi =\psi _{1}(x,y,z,t)\psi _{2}(\alpha
,\beta ,\gamma )$. Then $\rho =\rho _{1}(x,y,z,t)\rho _{2}(\alpha ,\beta
,\gamma )$, $S=S_{1}(x,y,z,t)+S_{2}(\alpha ,\beta ,\gamma )$ and
\begin{equation}
\Phi =\textstyle\int_{A}\bm{j}\cdot \bm{n}dA\textstyle\int \rho _{2}(\alpha
,\beta ,\gamma )d\mu (\alpha ,\beta ,\gamma ),  \label{eq:fluxA}
\end{equation}%
where $\bm{j}=\rho _{1}(x,y,z,t)\nabla S_{1}$ and $d\mu (\alpha ,\beta
,\gamma )=\sin \beta d\alpha d\beta d\gamma $. The particle flux $\Phi $ is
the only quantity directly accessible to the detector and depends only on
the spacetime part $\psi _{1}(x,y,z,t)$ of the wavefunction. As shown in
Eq.~(\ref{eq:fluxA}), the Euler's angles are integrated away for the simple
reason that the detector is located in the physical space and it is
insensitive to the particle orientation. It is worth noting that the
current density $j^{i}$ and, hence, the flux $\Phi $ is
Weyl-gauge invariant as it must be for any quantity having a measurable
value.\newline
Let us consider now the role played by the filtering apparatus. Unlike the
detector, whose role is just to count particles, the filtering stage of the
experimental setup must be tailored on the quantity to be measured. In the
case of the SGA, the filtering device is the spatial orientation of the
inhomogeneous magnetic field crossed by the particle beam. In an ideal
filtering apparatus no particle is lost, so its action on the particle's
wavefunction is "unitary". The role of the filter is to correlate the
spacetime path of the particle with the quantity to be measured (the spin
component, in our case) so to extract from the incident beam all particles
with a given value of the quantity (spin up, for example) by addressing them
to the appropriate detector. The filter acts on the particle motion in
space-time only. But, as said before, there is a feedback between the
particle motion and the geometric curvature of the configuration space, so
that the insertion of the filter changes not only the particle path in
spacetime, but also the overall geometry of the particle configuration
space, because it modifies its Weyl's curvature $R$. A similar mechanism is
at the core of General Relativity: the change in the motion, and/or the
addition of a massive body, changes the geometry of the whole surrounding
space. In our present approach, both particle motion and space geometry are
encoded in the scalar wavefunction, which indeed changes under the action of
the \textquotedblleft unitary\textquotedblright , i.e. lossless,
transformation introduced by the SGA filter. Solving the full dynamical and
geometric problem inside the SGA is a difficult problem, but the asymptotic
behavior of the scalar wavefunction far from the SGA is easily found. In
this \textquotedblleft far-field scattering approximation\textquotedblright
, a uniformly polarized particle beam is transformed by a SGA rotated at
angle $\theta $ with respect to the $z$-axis as follows,
\begin{eqnarray}
&&[aD_{\uparrow }(\alpha ,\beta ,\gamma )+bD_{\downarrow }(\alpha ,\beta
,\gamma )]\psi (\bm r,t)\overset{SGA}{\longrightarrow }  \nonumber
\label{eq:SGA} \\
&&\left( a\cos \frac{\theta }{2}+b\sin \frac{\theta }{2}\right) \!\!\left(
D_{\uparrow }(\alpha ,\beta ,\gamma )\cos \frac{\theta }{2}+D_{\downarrow
}(\alpha ,\beta ,\gamma )\sin \frac{\theta }{2}\right) \!\!\psi (\bm %
r_{u},t)+  \nonumber \\
&&\mbox{}+\left( a\sin \frac{\theta }{2}-b\cos \frac{\theta }{2}\right)
\!\!\left( D_{\uparrow }(\alpha ,\beta ,\gamma )\sin \frac{\theta }{2}%
-D_{\downarrow }(\alpha ,\beta ,\gamma )\cos \frac{\theta }{2}\right)
\!\!\psi (\bm r_{d},t)
\end{eqnarray}%
where $a,b$ are arbitrary complex constants with $|a|^{2}+|b|^{2}=1$, and
labels \textquotedblleft u\textquotedblright and \textquotedblleft
d\textquotedblright refer to the positions of the detectors located to the
up and down exit channels of the $\theta $-oriented SGA. The experimental
apparatus is arranged so that the wave packets $\psi (\bm r_{u},t)$ and $%
\psi (\bm r_{d},t)$ have negligible superposition and each detector sees a
wavefunction with space and angular parts factorized. Thus, for example, the
particle flux detected in the \textquotedblleft up\textquotedblright\
channel of the SGA is given, according to Eq.~(\ref{eq:fluxA}), by $\Phi
_{u}P_{u}(\theta )$, where $\Phi _{u}$ is the particle flux on the detector
and $P_{u}(\theta )=\left\vert a\cos \frac{\theta }{2}+b\sin \frac{\theta }{2%
}\right\vert ^{2}$ is usually interpreted as the probability that the
particle in the input wavepacket is found with its spin along the
\textquotedblleft up\textquotedblright direction of the SGA. As said above,
what the filter does is to correlate the particle space-time trajectory with
the quantity to be measured. In the standard quantum mechanical language, we
may say that the filter introduces a controlled entanglement among the
quantity to be measured and the particle spacetime path (in the SGA case,
the spacetime degrees of freedom become entangled with the orientational
ones). However, the filter is configured so that the wavepackets arriving on
each detector ($D_{u}$ and $D_{d}$, in our case) are not superimposed, and
the (approximate) wavefunction seen by each detector has the product form
considered above in Eq.~(\ref{eq:fluxA}). The last requirement ensures that
the detected particle flux $\Phi $ provides a correct measure (in the
quantum sense) of the measured quantity\footnote{%
It is precisely the lack of this condition which prevents to use the SGA to
measure the spin of electrons. A way to overcome this fundamental limitation
was proposed very recently~\cite{karimi12}.}.
\section{The EPR state and Bell inequalities}
Two equal\textit{\ }spin-1/2\textit{\ }particles $A$ and $B$, e.g. two
neutrons, propagate in opposite directions along the spatial $y$-axis $(%
\overrightarrow{y})$ of the Laboratory with a velocity $v\ll c$ towards two
spatially separate measurement devices, dubbed Alice and Bob, who measure
the spin of $A$ and $B$, respectively. Each apparatus, measuring the
particle $A$ (or $B$), consists of a standard Stern-Gerlach (SGA) device
followed by a couple of particle detectors that, being rigidly connected to
SGA, can \ be oriented with it by a rotation in the $\overrightarrow{x}$-$%
\overrightarrow{z}\ $plane at the corresponding angles$\ \theta _{A}$ (or $%
\theta _{B}$) taken respect to $\overrightarrow{z}$. Accordingly, $%
\overrightarrow{\theta _{A}}$ and $\overrightarrow{\theta _{B}}$ denote the
orientation axes of SGA$_{A}$ and SGA$_{B}$ \cite{Redhead}.

\bigskip Let's now turn our attention to the joint spin measurements of the
EPR entangled particles $A$ and $B$ described by Eq.~(\ref{eq:psiEPR}).
After leaving the source, particles $A$ and $B$ travel towards two Stern
Gerlach apparata, SGA$_{A}$ and SGA$_{B}$, respectively, located at Alice's
and Bob's stations on two distant sites along the $y$-axis. As said before,
each SGA acts \textit{locally}, by a \textit{unitary} transformation, on the
particle spatial, i.e. \textit{external}, degrees of freedom by correlating
its exit direction of motion with the direction of its spin respect to the
SGA axis, rotated around the $y$-axis at angle $\theta $, taken respect to
the $z$-axis. Since we are dealing with $\frac{1}{2}$-spins, there are only
two exit directions, either \textquotedblleft up\textquotedblright or
\textquotedblleft down\textquotedblright available to each particle which
will be then finally registered by a corresponding detector. Let's refer to
the Alice's and Bob's detectors as $D_{Au}$, $D_{Ad}$, $D_{Bu}$, $D_{Bd}$
and let $\theta _{A}$ and $\theta _{B}$ the angles of SGA$_{A}$ and SGA$_{B}$%
, respectively. Labels \textquotedblleft u\textquotedblright or
\textquotedblleft d\textquotedblright refer to the particle's exit
directions from each SGA's. As said above, the presence of the two SGA
changes not only the trajectories of the two particles, but also the Weyl
curvature of their configuration space. These changes are both encoded in
the change of the wavefunction $\psi _{AB}$ in Eq.~(\ref{eq:psiEPR}). Near
the source that wavefunction remains approximately unchanged, but far beyond
the spatial positions of the two SGA's the paths of the particles acquire
different direction according to their spin so that near the locations of
the detectors the input wavefunction is transformed according to
\begin{eqnarray}
&&\psi _{AB}\overset{SGAs}{\longrightarrow }A_{u,u}\psi _{A}(\bm{r}%
_{Au}t)\psi _{B}(\bm{r}_{Bu},t)+A_{u,d}\psi _{A}(\bm{r}_{Au},t)\psi _{B}(%
\bm{r}_{Bd},t)+  \nonumber \\
&&\mbox{}+A_{d,u}\psi _{A}(\bm{r}_{Ad},t)\psi _{B}(\bm{r}_{Bu},t)+A_{d,d}%
\psi _{A}(\bm{r}_{Ad},t)\psi _{B}(\bm{r}_{Bd},t)
\end{eqnarray}%
where $\bm{r}_{Au}$, $\bm{r}_{Ad}$, $\bm{r}_{Bu}$, $\bm{r}_{Bd}$ are the
positions of the detectors and $A_{u,u}$, $A_{u,d}$, $A_{d,u}$, $A_{d,d}$
are coefficients depending on the two particle Euler's angles and on the
angles $\theta _{A}$ and $\theta _{B}$ of SGA$_{A}$ and SGA $_{B}$,
respectively. The coefficients $A_{ij}$ $(i,j=u,d)$ can be easily calculated
by applying Eq.~(\ref{eq:SGA}):
\begin{subequations}\label{eq:Aij}
\begin{eqnarray}
A_{u,u} &=&\left( D_{\uparrow }(\alpha _{1},\beta _{1},\gamma _{1})\cos
\frac{\theta _{A}}{2}+D_{\downarrow }(\alpha _{1},\beta _{1},\gamma
_{1})\sin \frac{\theta _{A}}{2}\right) \times  \nonumber  \label{eq:A} \\
&&\times \left( D_{\uparrow }(\alpha _{2},\beta _{2},\gamma _{2})\cos \frac{%
\theta _{B}}{2}+D_{\downarrow }(\alpha _{2},\beta _{2},\gamma _{2})\sin
\frac{\theta _{B}}{2}\right) \sin \Delta \vartheta \\
A_{u,d} &=&\left( D_{\uparrow }(\alpha _{1},\beta _{1},\gamma _{1})\cos
\frac{\theta _{A}}{2}+D_{\downarrow }(\alpha _{1},\beta _{1},\gamma
_{1})\sin \frac{\theta _{A}}{2}\right) \times  \nonumber \\
&&\times \left( -D_{\uparrow }(\alpha _{2},\beta _{2},\gamma _{2})\sin \frac{%
\theta _{B}}{2}+D_{\downarrow }(\alpha _{2},\beta _{2},\gamma _{2})\cos
\frac{\theta _{B}}{2}\right) \cos \Delta \vartheta \\
A_{d,u} &=&\left( -D_{\uparrow }(\alpha _{1},\beta _{1},\gamma _{1})\sin
\frac{\theta _{A}}{2}+D_{\downarrow }(\alpha _{1},\beta _{1},\gamma
_{1})\cos \frac{\theta _{A}}{2}\right) \times  \nonumber \\
&&\times \left( D_{\uparrow }(\alpha _{2},\beta _{2},\gamma _{2})\cos \frac{%
\theta _{B}}{2}+D_{\downarrow }(\alpha _{2},\beta _{2},\gamma _{2})\sin
\frac{\theta _{B}}{2}\right) \cos \Delta \vartheta \\
A_{d,d} &=&\left( -D_{\uparrow }(\alpha _{1},\beta _{1},\gamma _{1})\sin
\frac{\theta _{A}}{2}+D_{\downarrow }(\alpha _{1},\beta _{1},\gamma
_{1})\cos \frac{\theta _{A}}{2}\right) \times  \nonumber \\
&&\times \left( -D_{\uparrow }(\alpha _{2},\beta _{2},\gamma _{2})\sin \frac{%
\theta _{B}}{2}+D_{\downarrow }(\alpha _{2},\beta _{2},\gamma _{2})\cos
\frac{\theta _{B}}{2}\right) \sin \Delta \vartheta
\end{eqnarray}%
where $\Delta \vartheta =\frac{1}{2}(\theta _{B}-\theta _{A})$.\newline

The coincidence rates are given by the joint particle fluxes intercepted by
the detectors
\end{subequations}
\begin{eqnarray}
\Phi _{i,j}(\theta _{A},\theta _{B}) &=&\textstyle\int \left\vert
A_{ij}(\alpha _{1},\beta _{1},\gamma _{1},\alpha _{2},\beta _{2},\gamma
_{2};\theta _{A},\theta _{B})\right\vert ^{2}d\mu (\alpha _{1},\beta
_{1},\gamma _{1})d\mu (\alpha _{2},\beta _{2},\gamma _{2})\times  \nonumber \\
&&\times \textstyle\int \bm{j}^{(A)}_{i}\cdot \bm{n}^{(A)}_{i}dA_{i}\textstyle\int \bm{j}%
^{(B)}_{j}\cdot \bm{n}^{(B)}_{j}dA_{j}\hspace{2cm}(i,j=u,d)
\end{eqnarray}%
and
\begin{equation}
\bm{j}^{(A,B)}_{i}=\left\vert \psi _{A,B}(\bm{r}_{i},t)\right\vert ^{2}\nabla S_{A,B}(%
\bm{r}_{i},t) \hspace{2cm}(i=u,d)
\end{equation}%
are the particle current densities at the detectors. A simple calculation
shows that if all particles falling into the detectors are counted, the
coincidence fluxes are given by
\begin{eqnarray}
\Phi _{u,u}(\theta _{A},\theta _{B}) &=&\Phi _{d,d}(\theta _{A},\theta _{B})=%
\frac{1}{2}\sin ^{2}\left( \Delta \vartheta \right) \\
\Phi _{u,d}(\theta _{A},\theta _{B}) &=&\Phi _{d,u}(\theta _{A},\theta _{B})=%
\frac{1}{2}\cos ^{2}\left( \Delta \vartheta \right) .
\end{eqnarray}%
in full agreement with standard quantum theory ~\cite{Bell,Scully}. This is
the key result of the present Article. The coincidence fluxes $\Phi _{ij}$
are Weyl-gauge-invariant and can be experimentally measured. Moreover, they
are equal to the joint probabilities $P_{i,j}(\theta _{A},\theta _{B})$
associated with the EPR state~(\ref{eq:psiEPR}) and lead straightforwardly
to the violation of Bell's inequalities within all appropriate experiments
consisting of statistical measurements over several choices of the angular
quantity $(\Delta \vartheta )$, as shown by many modern texts~\cite%
{Bell,Scully,Redhead}. For instance, Michael Redhead considers the
inequality:
\begin{equation*}
F(\Delta \vartheta )=\left\vert 1+2\cos (2\Delta \vartheta )-\cos (4\Delta
\vartheta )\right\vert \leq 2
\end{equation*}%
which is violated for all values of $\Delta \vartheta $ between $0$ and $%
45^{\circ }$ in a simple experiment~\cite{Redhead}.
\section{The no-signalling theorem}
Let us consider the EPR experiment discussed above as a paradigmatic example
of quantum nonlocality. The fluxes $\Phi _{ij}$ given in the preceding
Section are then identified with the probabilities of having the particle spins $%
(s_{A},s_{B})$ oriented in the directions $(i,j)$, respectively, with $%
i,j=(\uparrow ,\downarrow )$ and the integrands $|A_{ij}(\lambda ;\theta
_{A},\theta _{B})|^{2}$ as the joint probability distributions $p_{\lambda
}(i,j|\theta _{A},\theta _{B})$ to find $s_{A}=i$ and $s_{B}=j$ conditioned
by Alice's and Bob's respective SGA\ \ settings $\theta _{A},\theta _{B}$
for fixed values of the internal variables $\lambda =\{\zeta _{A}^{a},\zeta
_{B}^{b}\}$ spanning all six Euler angles of the two particles A and B. From
Eqs.~(\ref{eq:Aij}) we see that $p_{\lambda }(i,j|\theta _{A},\theta
_{B})=|A_{ij}(\lambda ;\theta _{A},\theta _{B})|^{2}$ have not the factor
form $p_{\lambda }(i|\theta _{A})p_{\lambda }(j|\theta _{B})$ required by
Bell's locality assumption. Hence, Quantum Geometrodynamics provides a
nonlocal (in Bell's sense)\footnote{%
Nonlocality is mathematically expressed by the dependence of $p_{\lambda
}(i,j|\theta _{A},\theta _{B})$ by Alice's and Bob's settings $\theta _{A}$
and $\theta _{B}$.} hidden variable completion to quantum mechanics, where
Bell's inequalities can be violated (and they are, indeed, as shown above)%
\footnote{%
Although not explicit demonstrated here, it is expected that this theory
would violate any other inequality based on Bell's locality assumption as,
for example, all inequalities~\cite{cabello08} used to prove the
Kochen-Specker theorem~\cite{kochen67}.}. But what about no-signalling?
Violating the no-signalling condition would led to serious problems against
the relativistic causality: information could be transferred between Alice
and Bob even if they were at space-like locations. No-signalling condition
is not required for the joint probabilities because there is no way to
transfer information exploiting spin correlations at Alice's and Bob's
sides. Quantum teleportation, for example, requires a classical
communication channel between Alice and Bob. No-signalling is required,
instead, for the marginal probabilities (and averages) measured at Alice and
Bob location. Mathematically, the no-signalling conditions are $p_{\lambda
}(i|\theta _{A},\theta _{B})=p_{\lambda }(i|\theta _{A})$ and $p_{\lambda
}(j|\theta _{A},\theta _{B})=p_{\lambda }(j|\theta _{B})$. These marginal
probabilities can be obtained from the joint probabilities $p_{\lambda
}(i,j|\theta _{A},\theta _{B})$ by summing out all variables which cannot be
controlled by the given observer. Then, the marginal probability densities
at Alice's and Bob's sides are given by
\begin{eqnarray}
p_{\lambda }(i|\theta _{A},\theta _{B}) &=&\int \sum_{j}|A_{ij}(\zeta
_{A}^{a},\zeta _{B}^{b};\theta _{A},\theta _{B})d\mu (\zeta _{B}^{b}) \\
p_{\lambda }(j|\theta _{A},\theta _{B}) &=&\int \sum_{i}|A_{ij}(\zeta
_{A}^{a},\zeta _{B}^{b};\theta _{A},\theta _{B})d\mu (\zeta _{A}^{a})
\end{eqnarray}%
respectively, where $\zeta _{A}^{a}=\{\alpha _{A},\beta _{A},\gamma _{A}\}$
and $\zeta _{B}^{b}=\{\alpha _{B},\beta _{B},\gamma _{B}\}$ and $d\mu (\zeta
)$ are the measures in the respective Euler's angle spaces. Now, a direct
calculation based on Eqs.(\ref{eq:Aij}) and on the explicit expression for
the normalized $D$-functions, yields~\footnote{%
When integrated over the respective Euler angles, the well-known quantum
result $P_{A}=P_{B}=\frac{1}{2}$ is obtained and Alice and Bob see their
respective particles fully unpolarized.}
\begin{eqnarray}
p_{\lambda }(i|\theta _{A},\theta _{B}) &=&\frac{1}{4}[1\pm (\cos \beta
_{A}\cos \theta _{A}+\cos \alpha _{A}\sin \beta _{A}\sin \theta _{A})]%
\hspace{3em}(i=\uparrow ,\downarrow )  \label{eq:pmarginal} \\
p_{\lambda }(j|\theta _{A},\theta _{B}) &=&\frac{1}{4}[1\pm (\cos \beta
_{B}\cos \theta _{B}+\cos \alpha _{B}\sin \beta _{B}\sin \theta _{B})].%
\hspace{3em}(j=\uparrow ,\downarrow )
\end{eqnarray}%
We see therefore that the marginal probability at each side of the EPR
experiment depends on the particle and SGA orientations at the same side
only. There is no way to Bob for sending signals to Alice by changing all he
can change: the angle $\theta _{B}$ of his SGA. The same is for Alice.
Quantum Geometrodynamics can then be considered as a nonlocal hidden
variables theory which, nevertheless, satisfies the no-signalling condition.
As final point we notice that the marginal probabilities (\ref{eq:pmarginal}%
) yields to marginal spin averages that are not of the scalar product form
required by Leggett's \textit{nonlocal} hidden variable models~\cite%
{leggett03}. Therefore, Quantum Geometrodynamics can violate -- and indeed
violates -- also Leggett's inequalities. 
\section{Conclusions}\label{sec:conclusions}
The above analysis shows that the "enigma" of quantum nonlocality, which is
generally considered to be epitomized by the violation of the Bell's
inequalities, may be understood on the basis of the Conformal Quantum
Geometrodynamics. This conclusion was obtained by a formal procedure
rigorous and exact, i.e. in absence of any approximation. The CQG theory
bears several appealing properties and may lead to far reaching consequences
in modern Physics. We summarize them as follows:

\begin{enumerate}
\item The linear structure of the standard first quantization theory is
fully preserved, in any formal detail and made compatible with the further
requirement of full conformal gauge invariance.

\item The conformal gauge can be chosen at will even if the equivalence
with standard quantum mechanics can be made only in one gauge where
the metric reduces to the Minkowski form. In this gauge gravitational and
quantum effects have different origin, namely the metric and the affine
connections, respectively. A gauge exists where quantum and
gravitational phenomena originate both from the metric tensor.

\item The quantum wavefunction acquires the precise meaning of a physical
quantum "\textit{Weyl's gauge field}" acting in a curved configurational
space. In particular, the square modulus of the wavefunction is identified
with the Weyl potential and its gradient with the Weyl vector.

\item A proper theoretical analysis of any quantum \textit{entanglement}
condition must involve the entire configurational space of the system
including the usual space-time of General Relativity as well as the
"internal coordinates" of the system. When entanglement is present and if
the internal coordinates are really "hidden", i.e. if they are absent in the
theory -- as they are generally considered in standard quantum mechanics --
severe limitations may arise on the actual interpretation of any dynamical
problem. The interpretation of physics may even be an impossible task, in
principle, and paradoxes may spring out. Indeed, in addition to "\textit{%
quantum nonlocality}", many counterintuitive concepts of quantum mechanics,
such as those related to several aspects of "\textit{quantum indeterminism}"
and of "\textit{quantum counterfactuality}" may arise from the theoretical
limitations due to the "incompleteness" of a description limited to
space-time fields. Which are indeed limitations to the human knowledge and
understanding.

\item The "sinister", "disconcerting" and "discomforting" aspects of \
entanglement were expressed right after the publication of the EPR paper by
a highly concerned Erwin Schr\H{o}dinger ~\cite{schrodinger35}. Who also added: "%
\textit{I would not call that one but rather the characteristic trait of
quantum mechanics, one that enforces the departure from the classical line
of thought}".
\end{enumerate}

We do believe that our present analysis enlightens from a novel insightful
perspective this highly intriguing aspect of modern Physics.

At last, and more generally, we believe that a quite interesting feature of
the present theory consists of its apparent unifying structure, connecting
for the first time General Relativity, Electromagnetism and Quantum
Mechanics within a unique (abelian) "gauge theory". It is also interesting,
not to say inspiring, to remark that a similar, non-abelian Yang-Mills gauge
theory underlies the electro-weak interactions and belongs to the "standard
model" of the elementary particles \cite{Quigg}.
%
\end{document}